\documentclass{aa}  
\usepackage{graphicx}
\usepackage{natbib}
\usepackage{txfonts}
\newdimen\captwidth
\newdimen\figwidth             
\newcommand{\gl}{\object{Gl~86}}
\newcommand{\rd}{\mathrm{d}}
\newcommand{\msun}{M_\odot}
\newcommand{\mj}{M_\mathrm{J}}
\newcommand{\ks}{K$_{\mathrm{s}}$}
\newcommand{\nds}{\mbox{ND}_\mathrm{short}}
\let\dy=\displaystyle
\newcommand{\excs}{\extracolsep{\fill}}
\begin{document}

\title{New constrains on Gliese 86 B\thanks{Based on ESO observing programs 70.C-0543, 072.C-0624 and 073.C-0468 at the VLT}}


\author{ A.-M. Lagrange\inst{1} 
\and H. Beust \inst{1}
\and S. Udry \inst{2}  
\and G. Chauvin \inst{3} 
\and M. Mayor \inst{2}}

\institute{Laboratoire d'Astrophysique de Grenoble, Universit\'e
J. Fourier, B.P. 53, F-38041 Grenoble Cedex 9, France\\
\email{lagrange@obs.ujf-grenoble.fr}
\and
Observatoire de Gen\`eve, 51 Ch. des Maillettes, 1290 Sauverny, Switzerland\\
\email{Stephane.Udry@obs.unige.ch}
 \and 
ESO Casilla 19001, Santiago 19, Chile\\
\email{gchauvin@eso.org}
}

\offprints{A.M. Lagrange} 

\date{Received ; Accepted}

 
  \abstract
   {}
   {We present the results of multi epochs imaging observations of the companion to the planetary host Gliese 86. Associated to radial velocity measurements, this study aimed at characterizing dynamically the orbital properties and the mass of this companion (here after Gliese 86 B), but also at investigating the possible history of this particular system.} 
   {We used the adaptive optics instrument NACO at the ESO Very Large Telescope to obtain deep coronographic imaging in order to determine new photometric and astrometric measurements of Gliese\,86\,B.} 
  {Part of the orbit is resolved. The photometry of \gl~B
  indicates colors compatible with a $\sim70$ Jupiter mass brown dwarf or a
  white dwarf. Both types of objects allow to fit the available, still
  limited astrometric data. Besides, if we attribute the long term
  radial velocity residual drift observed for \gl~A to B,
  then the mass of the latter object is $\simeq0.5\,M_\odot$. We analyse both astrometric and 
  radial velocity data to propose first orbital parameters for \gl~B.
  Assuming \gl~B is a $\simeq0.5\,M_\odot$ white dwarf, we explore
  the constraints induced by this hypothesis and refine the parameters
  of the system.}
   {}


   \maketitle
%

\section{Introduction}

One of the biggest challenges of today astronomy is to detect and
characterize extra solar planetary systems, and to understand the
way(s) they form and evolve. Over the past decade, the technical
improvements have allowed detections of more than 150 extrasolar
planets via radial velocity (hereafter RV) measurements down to 7.5
Earth Masses \citep{riv05} around
solar type stars, while direct imaging allows now the detection of
giant planets around young stars \citep{lagrange2004,chauvin2004}.
From the theoretical point of view the influence
of the multiplicity or companionship with outer bodies (e.g. brown
dwarfs; hereafter BD) on the dynamics and orbital stability of the
inner planets has been highlighted. This has led to constant efforts
trying to identify outer companions to those stars hosting planets
plus long term RV drifts.
\begin{table*}[htb]
\caption{Observation log. $\nds$ is a CONICA neutral
density filter with a transmission of 1.4\%. S13 and S27 are two
CONICA cameras corresponding respectively to a platescale of 13.25 and
27.01~mas. WFS corresponds to the wave front sensor of the adaptive
optics system.}
\label{table:1}
\begin{tabular*}{\textwidth}{@{\excs}llllllll}
\hline \noalign{\smallskip} 
UT Date & Filter & Camera & Observation
type & Exp. Time (s) & WFS & Obs-Program & Platescale calibrator\\
\noalign{\smallskip} \hline \noalign{\smallskip}
12/11/2003 & \ks & S27 & coronagraphy $(0.7~\!'')$ & $100\times0.6$ &
VIS & 072.C-0624 & $\Theta_1$ Ori C\\ 12/11/2003 & $2.17 +
\mbox{ND}_{\mathrm{short}}$ & S27 & direct & $15\times4.0$ & VIS
&072.C-0624 & $\Theta_1$ Ori C\\ \noalign{\smallskip}
22/09/2004&H & S13 & coronagraphy $(0.7~\!'')$ & $48\times1.0$ & VIS &
073.C-0468 &$\Theta_1$ Ori C \\ 22/09/2004&H + $\nds$ &
S13 & direct & $42\times0.35$ & VIS & 073.C-0468 &$\Theta_1$ Ori C \\
\noalign{\smallskip} 
29/07/2005 & \ks & S27 & coronagraphy $(0.7~\!'')$
&$400\times0.8$ & VIS & 075.C-0813 & $\Theta_1$ Ori C\\ 
29/07/2005 & \ks + $\nds$ & S27 &direct &$400\times0.35$ & VIS &
075.C-0813 & $\Theta_1$ Ori C\\
29/07/2005 & H & S13 & coronagraphy
$(0.7~\!'')$ &$360\times1.$ & VIS & 075.C-0813 & $\Theta_1$ Ori C\\
29/07/2005 &H + $\nds$ & S13 &direct &$400\times0.35$ &
VIS & 075.C-0813 & $\Theta_1$ Ori C\\ 
29/07/2005 & J & S13 &
coronagraphy $(0.7~\!'')$ &$165\times2.$ & VIS & 075.C-0813 &
$\Theta_1$ Ori C\\
29/07/2005 & J + $\nds$ & S13 &direct
&$240\times0.5$ & VIS & 075.C-0813 & $\Theta_1$ Ori C\\
\noalign{\smallskip} \hline
\end{tabular*}
\label{logobs}
\end{table*}

\gl~A is a K0V star with an estimated mass of $0.8\,\msun$
\citep{siess1997,baraffe1998} and is located at $10.9$\,pc from the
Sun \citep{perryman1997}. Through RV measurements, \citet{queloz2000}
have detected a 4~$\mj$ (minimum mass) planet \gl~b, orbiting \gl~A at
$\sim 0.11\mbox{AU}$. This star is also surrounded by a more distant
companion \gl\,B, discovered at $\sim 20\,\mbox{AU}$ using
coronagraphy coupled to adaptive optics imaging \citep{els2001}. The
estimated photometry of \gl~B is compatible with that expected for a
$40$--$70\,\mj$ brown dwarf companion. Howewer, \citet{mug05}
showed recently that this was also compatible with a cool white dwarf,
and that the latter hypothesis was more likely regarding the K band
spectrum of the companion. The absence of near-IR molecular and atomic
lines as well as the steep K-band continuum are indeed consistent with
what is expected for a high gravity object with an effective
temperature higher than 4000~K.

Apart for the RV wobble due to the hot Jupiter companion,
\gl~A also exhibits a long term RV drift measured with \textsc{Coravel} 
and \textsc{Coralie} over 20 years.
This drift indicates the possible presence of an additional more distant
companion, with a
substellar mass and a distance to star greater
than $\simeq$ 20 AUs. \citet{els2001} claimed that \gl~B cannot
account for this RV drift, due to its too low mass.
They postulated instead that an additional
companion, located in 2000 ``behind'' the star (i.e., under the coronographic
mask), could be responsible for the observed drift.

In the course of a deep search for faint outer companions to stars
hosting planets with NACO, we were able to make new images of \gl~A
and B in the near IR. We present the observational results in
Sect.~2. In Sect.~3, we report new photometric result of \gl~B and we
present an analysis of both astrometric and RV data, assuming that the
RV drift is due to \gl~B. Finally, in Sect.~4 we discuss the nature of
\gl~B, and we confirm that it is very probably a $\sim 0.5\,\msun$ white
dwarf. We discuss the implications of this hypothesis.
\section{Observations}
\subsection{NACO observing log}
\begin{figure}[t] 
\centering \includegraphics[width=7cm]{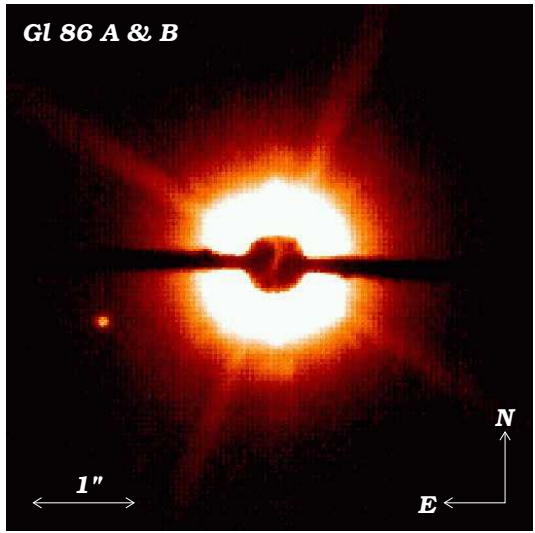}
\caption{VLT/NACO \ks-band coronagraphic image of \gl~A and B,
acquired on September 24, 2004, with an occulting mask of diameter 0.7\arcsec.}
\label{obs1}
\end{figure}
Observations of \gl\ were performed on November 12, 2003, September 22,
2004 and uly 29, 2005 with NACO at the VLT. NACO is equipped with an
adaptive optics system \citep{rousset2002,lagrange2002} that provides
diffraction limited images in the near infrared (IR) and feeds the
observing camera CONICA \citep{lenzen2002}. Both coronagraphic and
direct images were performed to image respectively \gl~B and A within
the linearity domain of the detector. Note that between the two
observing dates, the CONICA detector was changed and the latter
detector was more efficient.

The calibrations of platescale and detector orientation were done
using the $\Theta_1$ Ori C astrometric field on November 12, 2003,
September 22, 2004 and July 29,2005. On November 12, 2003 and July 29,
2005, the orientation of true north of the S27 camera was found
respectively at $-0.06\degr$ and $-0.05\degr$ east of the vertical with an
uncertainty of $0.20\degr$ and the platescale was $27.01\pm0.10$\,mas. On
September 22, 2004, the orientation of true north of the S13 camera was
found $0.20\degr$ east of the vertical with an uncertainty of $0.20\degr$
and the platescale was $13.25\pm0.10$\,mas.
Table~\ref{logobs} summarizes the new observations as well as archival
ones and Fig.~\ref{obs1} shows a \ks\ ($\lambda_{\mathrm{c}}=2.2\,\mu$m,
$\Delta_{\lambda}=0.35\,\mu$m) image recorded in September 2004.
\subsection{Photometric measurements}
\begin{table}[b]
\caption{Photometry of Gl86\,A and B}
\label{table:2}
\centering
\begin{tabular*}{\columnwidth}{@{\excs}llll}     
\hline\hline\noalign{\smallskip} 
Component         &  J  & H    &  K          \\
    &  (mag)         &  (mag)    & (mag)               \\
\noalign{\smallskip} \hline\noalign{\smallskip} 
Gl86\,A$^{a}$ & $4.79 \pm 0.03$ & $4.25 \pm 0.03$ & $4.13 \pm 0.03$      \\
Gl86\,B$^{b}$ & $14.7 \pm 0.2$ & $14.4 \pm 0.2$ & $13.7 \pm 0.2$     \\
Gl86\,B$^{c}$ & $12.9 \pm 0.3$ & $13.1 \pm 0.2$ & $12.8 \pm 0.2$     \\
\noalign{\smallskip} \hline
\end{tabular*}
\begin{list}{}{}
\item[$^{\mathrm{a}}$] from the 2MASS All-Sky Catalog \citep{cutr03}.
\item[$^{\mathrm{b}}$] from \citet{els2001}
\item[$^{\mathrm{c}}$] from $^{\mathrm{a}}$ and NACO measurements
presented in this work.
\end{list}
\end{table}
\begin{figure*}[t]
\makebox[\textwidth]{
\includegraphics[angle=-90,origin=br,width=\columnwidth]{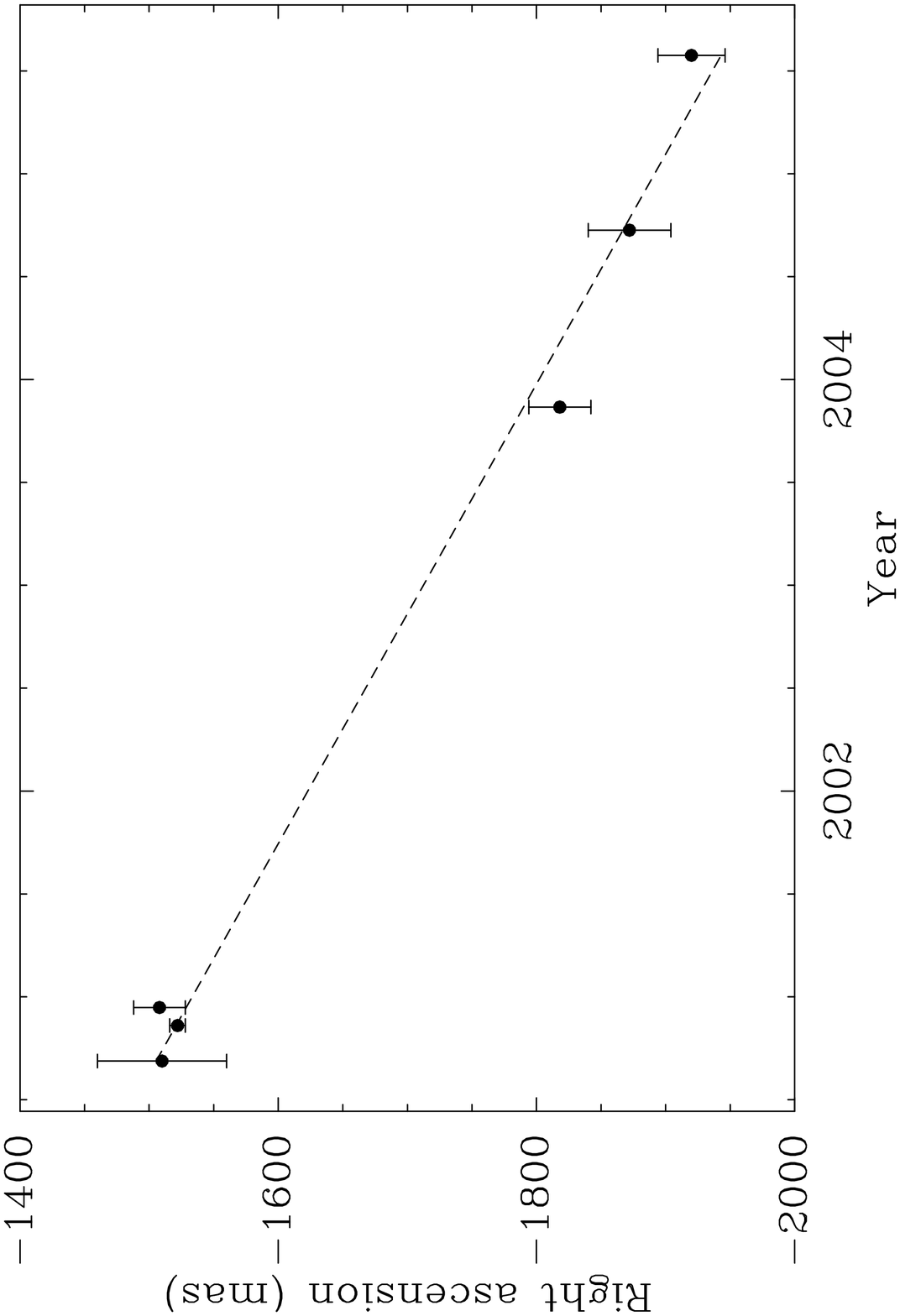}
\hfil
\includegraphics[angle=-90,origin=br,width=\columnwidth]{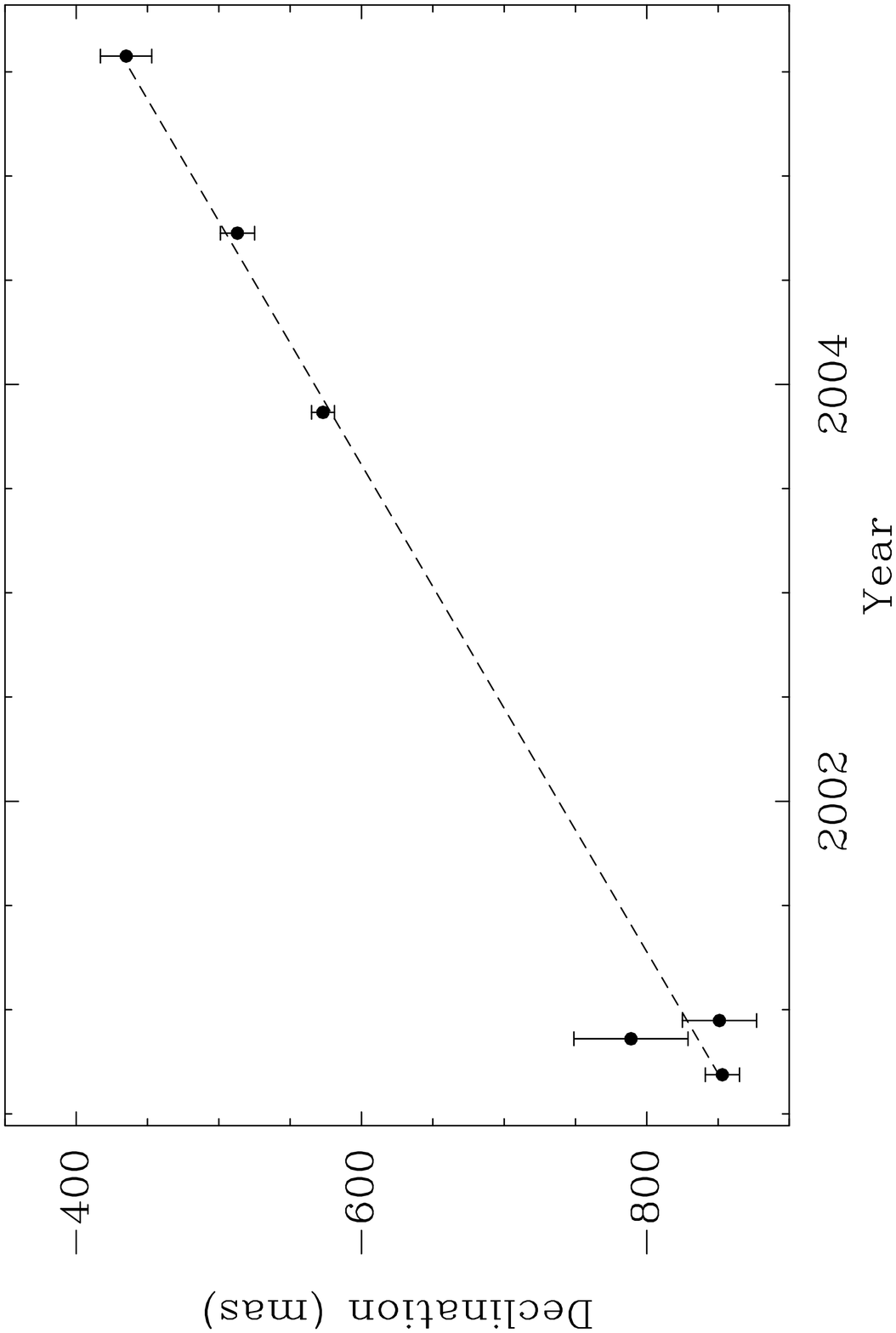}}
\caption[]{Least square fits of the right ascension (left)
and of the declination (right) of \gl~B relative to A.}
\label{mvtfit}
\end{figure*}
On July 29, 2005, the NACO measurements of Gl86\,A and B were obtained
in J,H and \ks\ filters, under photometric conditions. The
JH\ks\ contrasts were determined using the deconvolution algorithm
of \citet{vera98}. Based on the JHK photometry of Gl86\,A
from the 2MASS All-Sky Catalog \citep{cutr03}, we then deduced
the JHK photometry of Gl86\,B (see Table~\ref{table:1}). The
transformation between the K$_{s}$ filter of NACO and the K filter
used by CTIO-2MASS was found to be smaller than 0.03 magnitude.

The reported photometry is significantly different from the ones given
by \cite{els2001} (see Table~\ref{table:1}). This could be related to
systematic photometric errors induced by an incorrect subtraction of
the \gl\,A PSF wings within the coronagraphic image. This effect is
generally more important for shorter wavelengths where AO corrections
is poorer. It often leads to an underestimation of the companion flux,
as this is the case in Table~\ref{table:1} when comparing ADONIS and
NACO data. Thanks to the highest angular resolution and the enhanced
detection capabilities provided by NACO at VLT, we can reasonably
expect to be less sensitive to this PSF subtraction effect to derive
the \gl\,B photometry.

The new NACO photometry found is still compatible with the conclusions
of \cite{els2001} that Gl86\,B has a photometry similar to that
expected for a substellar companion with a mass of $40$--$70\,\mj$
(spectral type L7--T5). However, this photometry can also correspond
to the one expected for a cool white dwarf and, as recently claimed by
\cite{mug05}, this is more likely the case as the spectrum of \gl~B
does not exhibit the molecular absorption features in K band that are
characteristic for L or T dwarfs. \\

In the following, we reinvestigate this issue (brown or white dwarf)
from a dynamical point of view.
\subsection{Astrometric measurements}
The offset positions of \gl~B to A, recorded with NACO on 12 November
2003, 22 September 2004 and 29 July 2005, were translated into
physical values using the corresponding astrometric calibration
data. The shifts induced by the use of different filters between
coronographic and direct images were taken into
account. Table~\ref{astromdata} summarizes the measured values and
Figure~\ref{mvtfit} shows the various data points in a
($\Delta\alpha$, $\Delta\delta$) diagram, as well as the offset
positions of \gl~B to A measured by \citep{els2001} with
ADONIS/SHARPII on 8 September 2000. The orbital motion of \gl~B is
clearly identified. This confirms the independent detection of
\cite{mug05}.
\begin{table}[t]
\caption[]{Offset positions of the \gl~B relative to A}
\label{astromdata}
\begin{tabular*}{\columnwidth}{@{\excs}llllll}
\hline\hline\noalign{\smallskip} 
UT Date  & Julian &$\Delta\alpha$  & $\Delta\delta$ & Separation & Position\\
& Date & (mas) &  (mas) &(mas)  & Angle ($\degr$)\\
\noalign{\smallskip}\hline\noalign{\smallskip}
08/09/2000 & 2451796 & $1510\pm25$   & $-853\pm6$ &$1734\pm22$&$119.5\pm0.8$\\
10/11/2000  & 2451859 & $1522\pm3$   & $-789\pm20$ &$1714\pm10$&$117.4\pm0.4$\\
12/12/2000  & 2551891 & $1508\pm10$   & $-851\pm13$ &$1732\pm11$& $119.4\pm0.4$\\
\noalign{\smallskip}
12/11/2003 & 2452986 & $1818\pm12$   & $-573\pm4$ &$1906\pm11$ & $107.5\pm0.4$\\
22/09/2004 & 2453271 & $1872\pm16$ & $-513\pm6$ & $1941\pm14$  & $105.3\pm0.5$\\
29/07/2005 & 2453581 & $1920\pm13$ & $-435\pm9$ & $1969\pm11$  & $102.7\pm0.4$\\
\noalign{\smallskip}\hline
\end{tabular*}
\end{table}
\subsection{Radial velocity data}
Radial velocity measurements of \gl~A have been
gathered for more than 20 years now.
The whole data set reveals in addition to a short period
modulation of $\sim 1\,\mbox{km\,s}^{-1}$ amplitude that has been
attributed to a hot Jupiter companion 
\citep{queloz2000}, the presence of a regular continuous
decrease of $\sim 2\,\mbox{km\,s}^{-1}$ in 25 years
(Fig.~\ref{gl86orbvrad}).

It is tempting to try to attribute this regular decrease to \gl~B.
The temporal derivative of the radial velocity of the primary in a binary
system is easy to derive. One gets
\begin{equation}
\frac{\rd v_\mathrm{r}}{\rd t}=-\frac{Gm}{r^2}\sin i\sin(\omega+v)
\end{equation}
where $G$ is the gravitational constant, $m$ is the mass of the companion
$r$ is the distance between the two bodies, $i$ is the inclination
of the orbit with respect to the plane of the sky, $\omega$ is the
argument of periastron, and $v$ is the true anomaly, i.e. the current
polar position along the orbit with respect to the periastron.
Of course most of these quantities are unknown, but a simple application
assuming $\sin i\sin(\omega+v)\simeq 0.5$ and $r\simeq 20\,$AU shows
that $\rd v_\mathrm{r}/\rd t=\sim 2\,\mbox{km\,s}^{-1}/25\,\mbox{yr}$
is hardly compatible with $m=70\,\mj$, but rather with $m$ ranging between
0.2 and $1\,\msun$.

This result led \cite{els2001} to conclude that the RV residuals
are not due to \gl~B, but rather to an unseen, additional body.
Conversely if we keep attributing the RV decrease to \gl~B, 
this raises the question of the mass of \gl~B.  The available photometry is
compatible with a $70\,\mj$ object \citep{els2001}. But it can also
be compatible with a $\sim 0.5\,\msun$ object if this object is a white
dwarf.
Obviously, more data, in particular spectroscopic data
are needed to discriminate between these two possibilities.
\section{Data Analysis}
\subsection{General analysis of astrometric data}
From Fig.~\ref{mvtfit}, one can see that 
on the plane of the sky, the four points 
(see plots below) are roughly aligned, so that the only relevant
information we can derive from these data is a middle astrometric
position (at $t=2003.0$) and temporal derivatives of the right ascension
$\alpha$
and of the declination $\delta$. We thus perform a least-square fit
of the available data to derive them. The result is shown on
Fig.~\ref{mvtfit}. We see that $\alpha$ and $\delta$ actually vary roughly
linearly with time. The linear fit is therefore relevant.
The corresponding temporal derivatives are
\begin{equation}
\left\{\begin{array}{rcl}
\dy\frac{\rd(\alpha)}{\rd t} & = & \dy
-89.5\pm 8.7\:\mathrm{mas}\,\mathrm{yr}^{-1}\\[2\jot]
\dy\frac{\rd(\delta)}{\rd t} & = & \dy85.6\pm 7.18\:
\mathrm{mas}\,\mathrm{yr}^{-1}\end{array}\right.
\end{equation}
These derivative values, together with the mean present values of
$\alpha$ and $\delta$, provide four contraints on the orbit of the
companion with respect to the primary. In principle, this orbit
is fully characterized by 6 orbital elements, plus the unknown
mass $m$ of the companion. The constraints allow us to fix 4 of them.
We chose to let the mass $m$ of the companion, the inclination
$i$ with respect to the plane of the sky, and the longitude
of the ascending node $\Omega$ as free parameters. For any given set
of parameters ($m,i,\Omega$) we are able to derive the remaining ones,
i.e. the semi-major axis $a$, the eccentricity $e$, the argument
of periastron $\omega$ and the mean anomaly $M$. We recall that $M$
is a quantity that characterizes the present position of the
companion on its orbit. $M$ is proportional to the time,
$M=0$ at periastron and $M=2\pi$ one orbital period later.
\subsection{Analysis assuming that \gl~B is a $70\,\mj$ object}
\begin{figure}
\includegraphics[angle=-90,origin=br,width=\columnwidth]{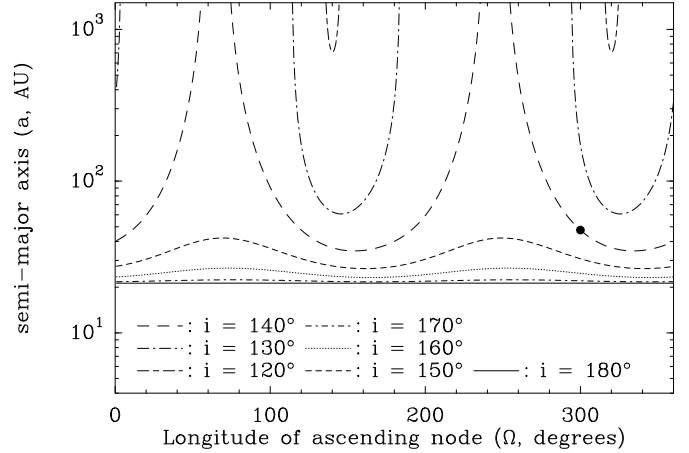}
\caption[]{The semi-major axis $a$ of the orbital solution for the
\gl\ companion, as a function of the longitude of the ascending
node $\Omega$, for various values of the inclination $i$ between
$120\degr$ and $180\degr$, for a fixed companion mass 
$m=70\,M_\mathrm{J}$. The bullet represents the solution plotted in
Fig.~\ref{gl86orb} (upper plot) and detailed in Eq.~(\ref{sol1}).}
\label{solm70a}
\end{figure}
\begin{figure}
\includegraphics[angle=-90,origin=br,width=\columnwidth]{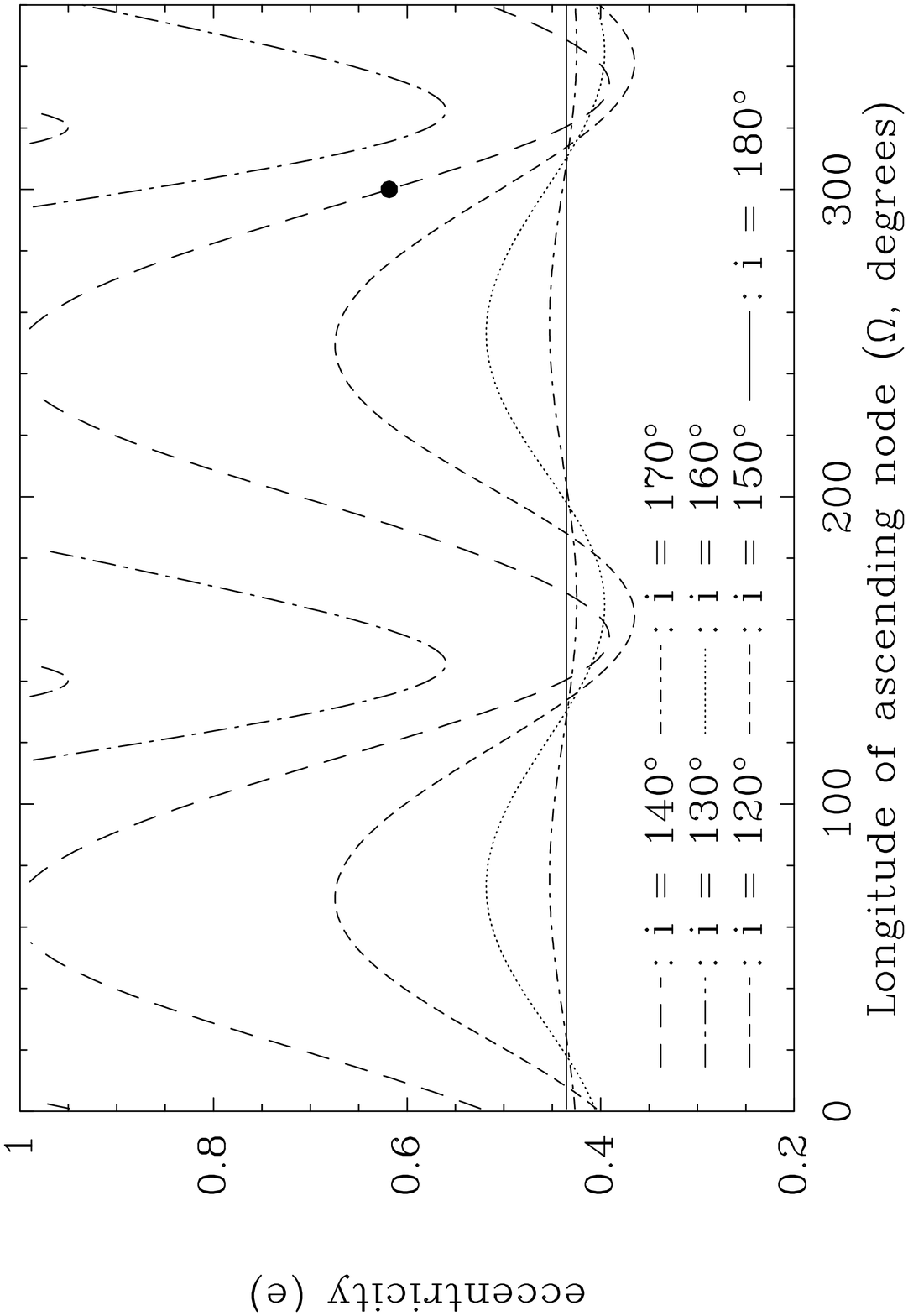}
\caption[]{Same as Fig.~\ref{solm70a}, but for the orbital eccentricty
of the solution}
\label{solm70e}
\end{figure}
\begin{figure}
\includegraphics[angle=-90,origin=br,width=\columnwidth]{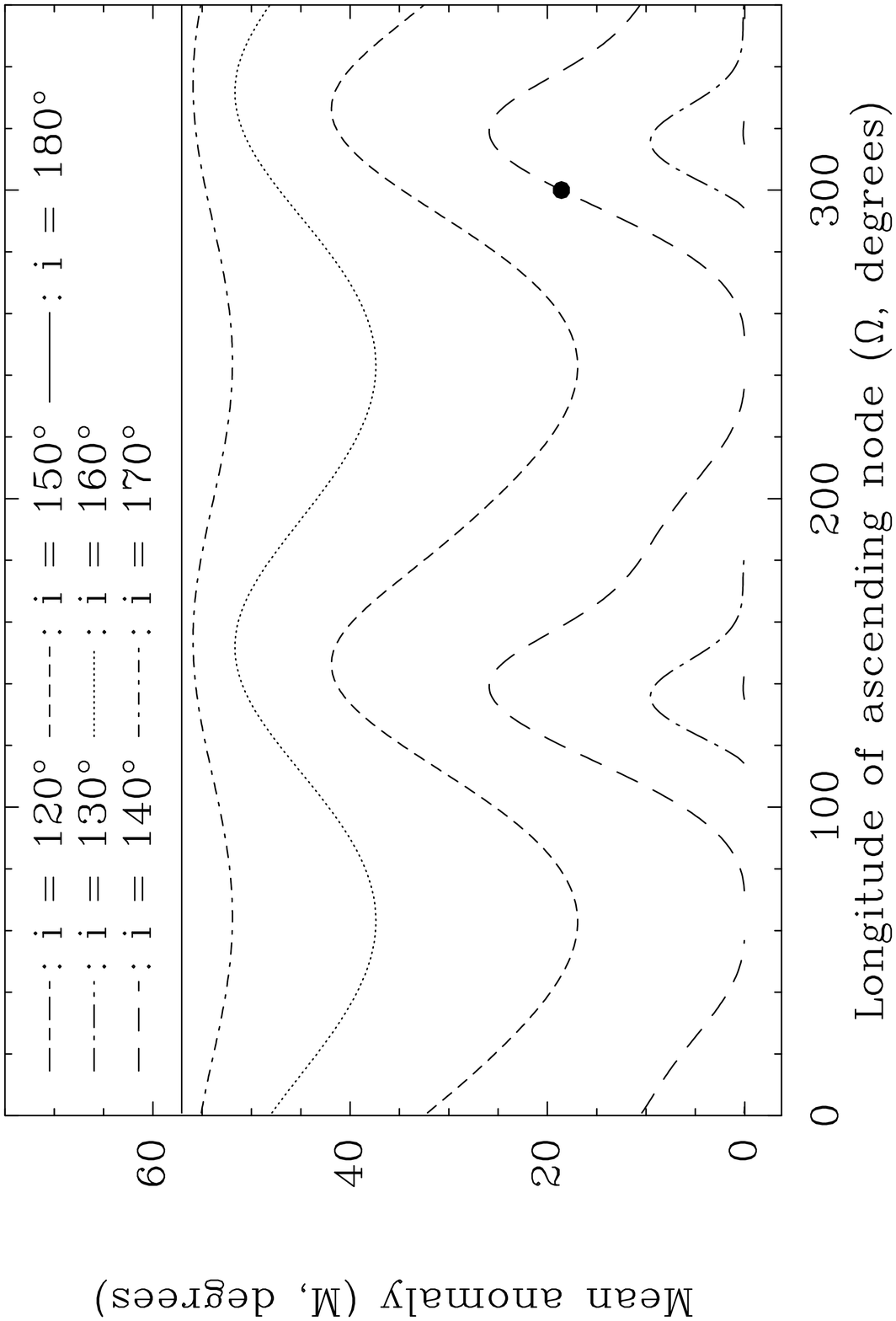}
\caption[]{Same as Fig.~\ref{solm70a}, but for the present mean anomaly $M$}
\label{solm70m}
\end{figure}
Depending on the free parameter set we choose, there is not necessarily
an orbital solution compatible with the constraints. In particular, it turns
out that there is no solution for $i<120\degr$. This means that we are
viewing the orbit nearly from its south pole. The result of the parameter
space exploration is shown in Figs.~\ref{solm70a}--\ref{solm70m}.
The semi-major axis, the eccentricity, and the mean anomaly are plotted
as a function of $\Omega$, for different values of the inclination $i$,
and for a fixed companion mass $m=70\,\mj$. We note that in some
cases ($i=120\degr$) there is not
a solution for every $\Omega$ value. We note also that the orbit is
necessarily eccentric ($e>0.35$ in any case), and that in all cases,
the companion is presently short after periastron ($0<m<60\degr$).
Of course we explored other companion masses in the compatible range
($60\,\mj\la m\la 90\,\mj$). The result is not
shown here but it is nearly equivalent to that for $m=70\,\mj$.
Actually Figs.~\ref{solm70a}--\ref{solm70m} represent the standard solution.
\begin{figure*}
\includegraphics[angle=-90,origin=br,width=\textwidth]{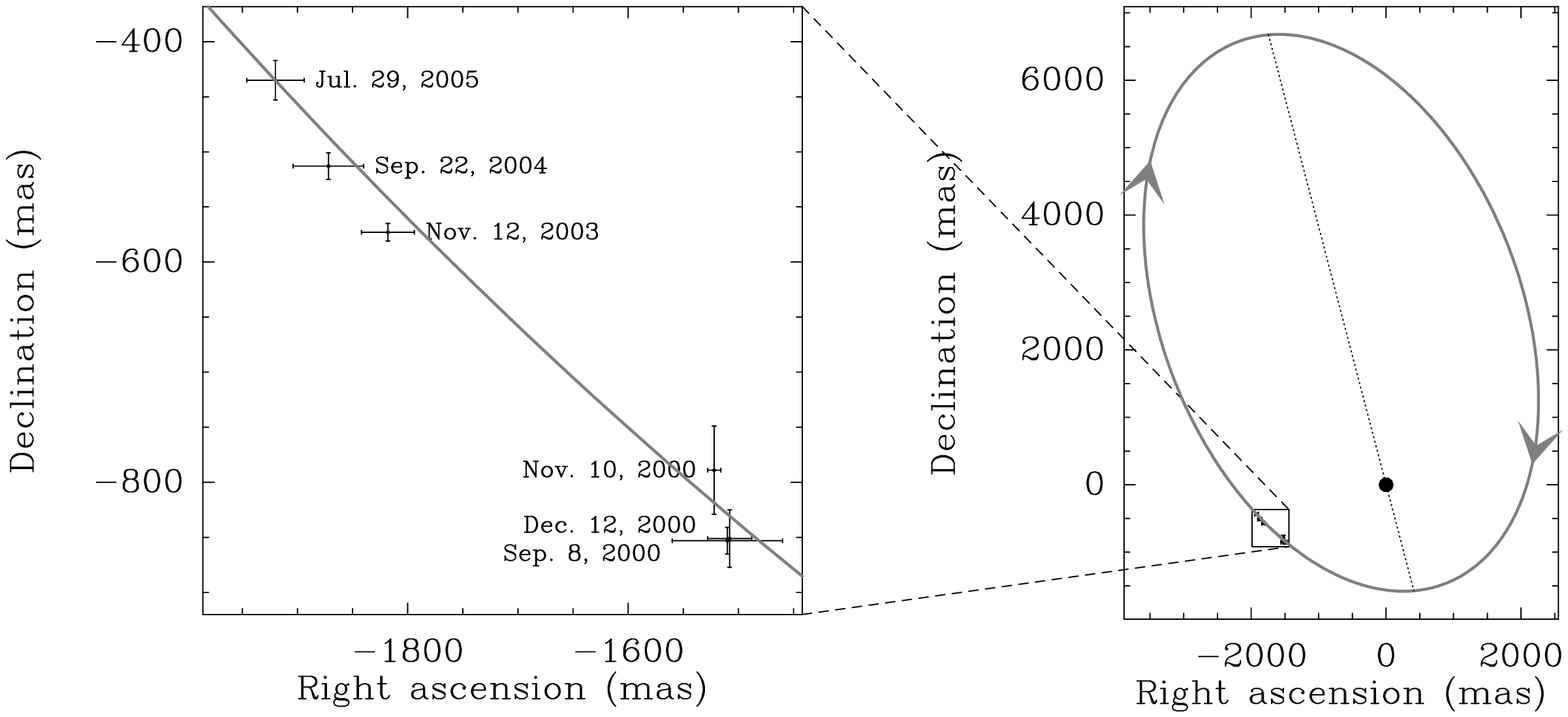}
\includegraphics[angle=-90,origin=br,width=\textwidth]{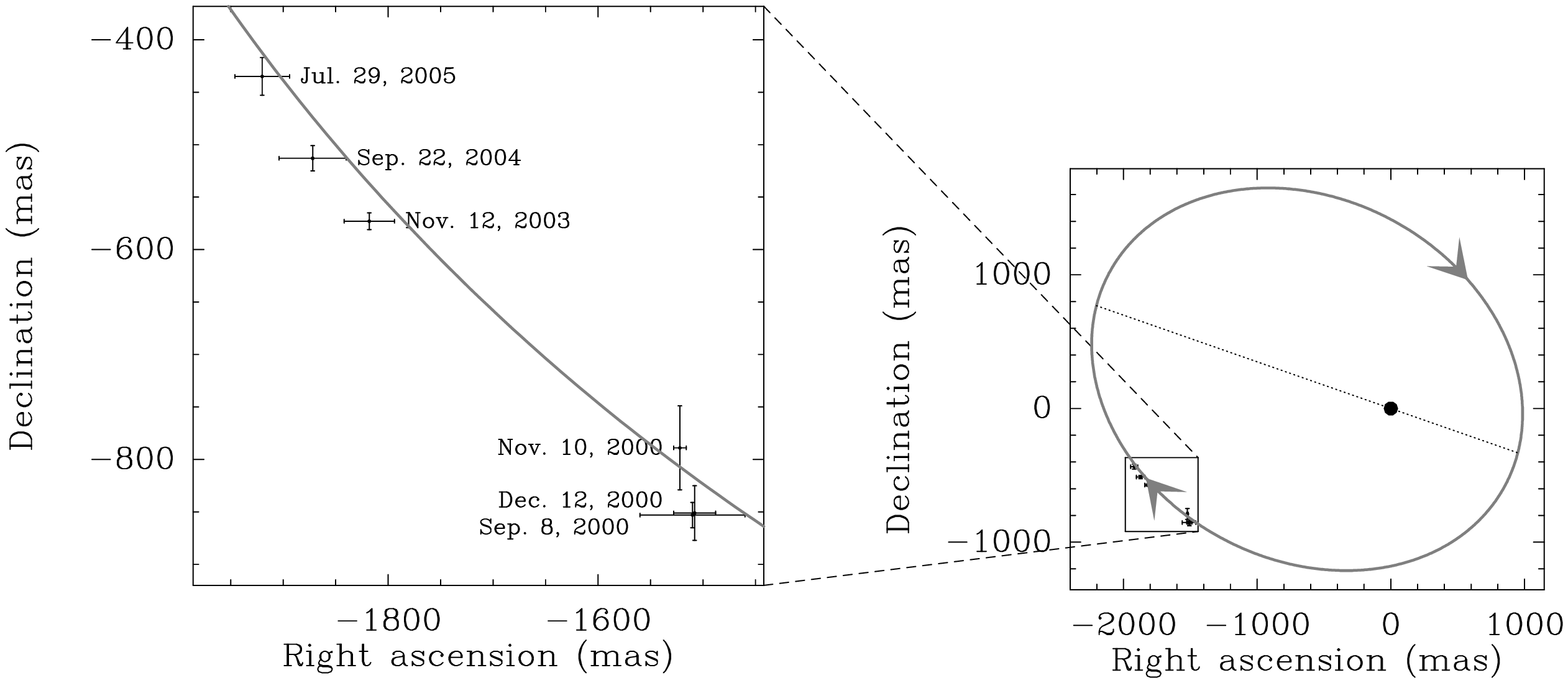}
\caption[]{A representation of the orbital solutions described by
Eq.~(\ref{sol1}) (upper plot, brown dwarf case) and 
Eq.~(\ref{solpart}) (lower plot, white dwarf case),
as projected onto the
plane of the sky. In each case, The right plot represents a view of
the full orbit
and the left plot is an enlargement of the present day motion.
The dotted line is the projection of the line of apsides of the orbit.}
\label{gl86orb}
\end{figure*}
In order to better show the shape of the orbital solution, we display
one particular, typical solution, marked as a bullet in
Figs.~\ref{solm70a}--\ref{solm70m}, and charaterized by $i=150\degr$,
and
\begin{equation}
\left\{
\begin{array}{l}
\dy a=47.58\,\mbox{AU}\;,\quad e=0.6185\;,\quad\Omega=300\degr\;,\\[\jot]
\dy \omega=19.71\degr\;,\quad M=18.58\degr
\end{array}\right.
\label{sol1}
\end{equation}
The projection of this solution onto the plane of the sky is shown 
in Fig.~\ref{gl86orb}. We clearly see that the orbit is eccentric
and that the present day position of the companion is short after 
periastron. The associated orbital period is 353\,yr, and the last
periastron passage turns out to have occurred in 1984. Of course
the latter quantities are subject to some variations if we consider another
solution.

In Fig.~\ref{gl86orbvrad}, we show the \gl\ radial velocity data set,
superimposed to the theoretical curve that would be expected for the
solutions we display in Fig.~\ref{gl86orb}. Note that in those curves,
we dot not add the short period modulation due to the hot Jupiter
companion, as this object produces a much smaller amplitude. 
We also add to the theoretical radial velocity curve an
empirical offset, intended to correspond to the mean heliocentric
velocity of the \gl\ system, fixed in such a way that the radial
velocity matches the mean observed value in 2003.0 . Actually
the only relevant parameter we need to compare between the data
and the theory is the mean temporal derivative of the radial velocity
at in 2003.0, and also the general trend over 25 years. 

In Fig.~\ref{gl86orbvrad}, the theoretical radial velocity 
curve corresponding to Eq.~(\ref{sol1}) is represented as a dashed line.
We see that it does not match the data. In fact the decrease in 2003.0 is only
10\%\ of the observed values ($0.1\,\mbox{km\,s}^{-1}/25\,\mbox{years}$).
As explained above, this was expected from our order of magnitude
estimate of the mass needed to account for the observe decrease rate.
\subsection{Analysis assuming \gl~B is a $\sim 0.5\,\msun$ object}
\label{fitwd}
\begin{figure}
\includegraphics[angle=-90,origin=br,width=\columnwidth]{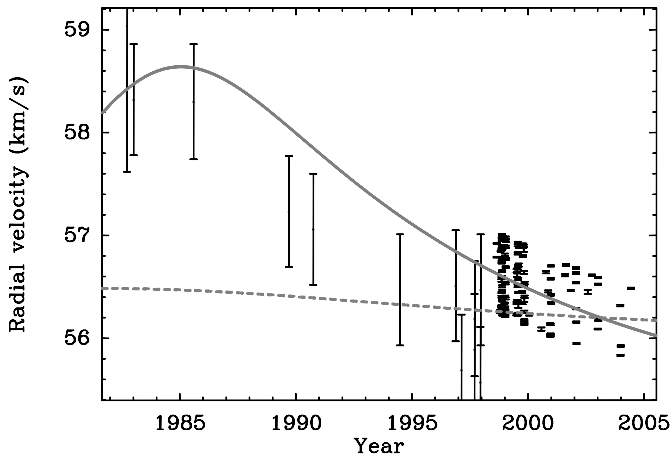}
\caption[]{Radial velocity data of \gl~A as monitored 
over 25 years,
superimposed to the theoretical curves (in grey) corresponding
to the orbital solutions displayed in Fig.~\ref{gl86orb}. The low
accuracy data up to 1998 are the \textsc{Coravel} data
(typical error $\pm0.27\,\mbox{km}\,\mbox{s}^{-1}$),
while the subsequent high accuracy data are the \textsc{Coralie}
ones (typical error $\pm0.005\,\mbox{km}\,\mbox{s}^{-1}$).
The dashed curve corresponds to the orbit detailed in Eq.~(\ref{sol1})
where \gl~B is taken as a brown dwarf. The solid curve corresponds
to the orbit described in Eq.~(\ref{solpart}), where \gl~B is fitted
as a white dwarf. The fit of the radial velocity residuals
is much better.}
\label{gl86orbvrad}
\end{figure}
If we now assume that the residuals of the radial velocity data are
due to \gl~B, we get additional constraints to the orbital
parameters. In particular, we can force the temporal derivative of the
radial velocity in 2003.0 to match the observed one. This in turn
enables to fix the mass $m$ of the companion instead of giving it as
an input parameter.  However, this single criterion turned out not to
be sufficient.  We may derive solutions that fit the radial velocity
derivative in 2003.0 but that do not fit the radial velocity data over
the whole observation period, especially the older data.  Hence we
retain in the fitted solutions only those which fit a convenient least
square criterion with the whole radial velocity data sample.

\begin{figure}
\includegraphics[angle=-90,origin=br,width=\columnwidth]{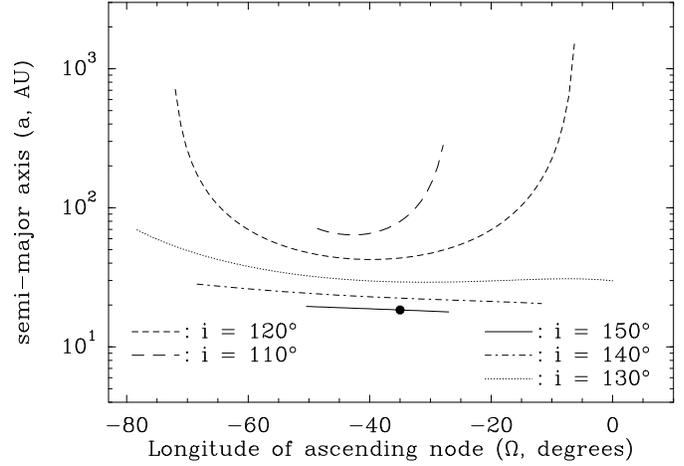}
\caption[]{The semi-major axis $a$ of the orbital solution for
\gl~B that also fits the radial velocity data residuals,
as a function of the longitude of the ascending
node $\Omega$, for various values of the inclination $i$ between
$110\degr$ and $150\degr$. The bullet represents the solution plotted in
Fig.~\ref{gl86orb} (lower plot), described in Eq.~(\ref{solpart}).}
\label{solva}
\end{figure}
\begin{figure}
\includegraphics[angle=-90,origin=br,width=\columnwidth]{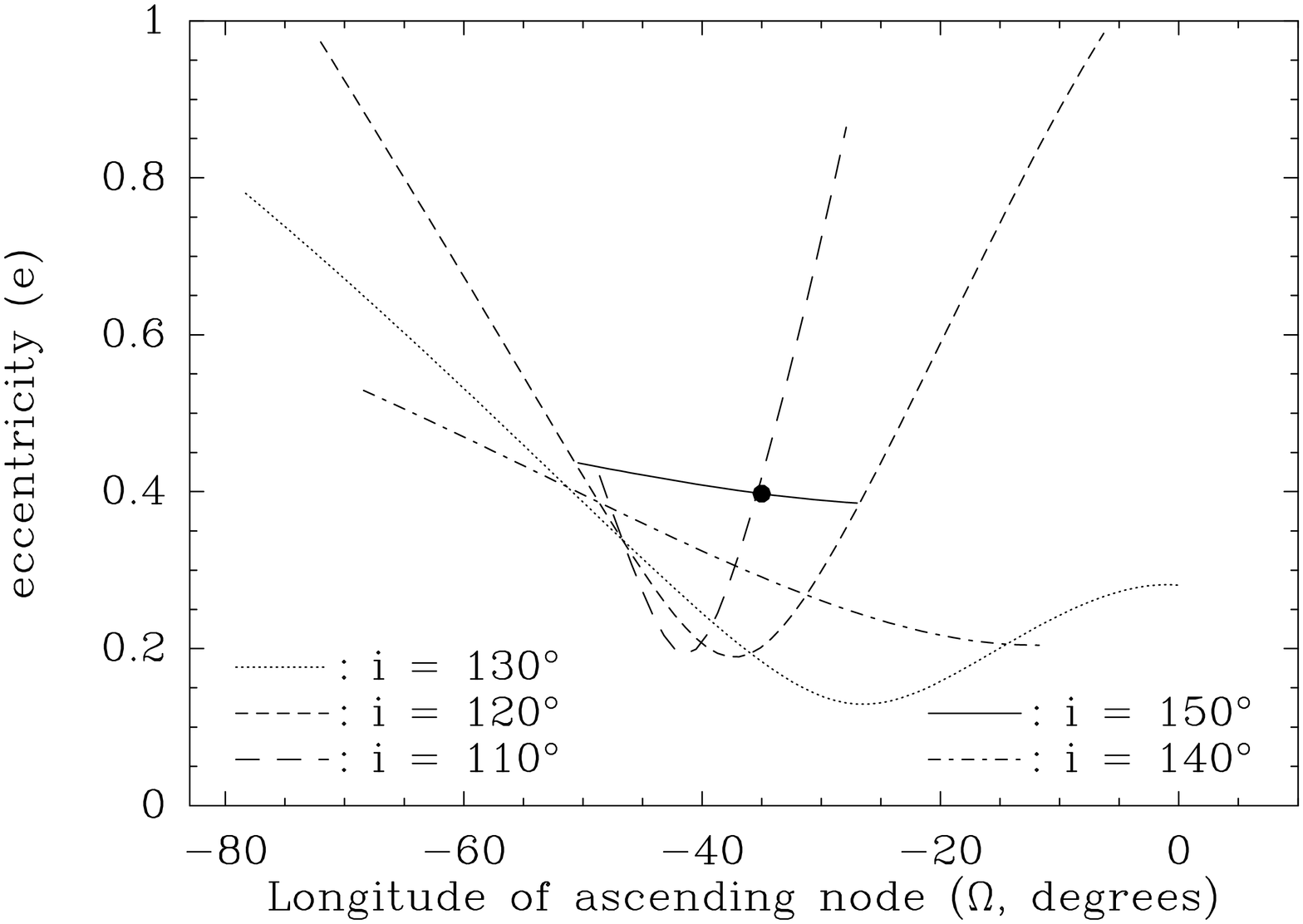}
\caption[]{Same as Fig.~\ref{solva}, but for the orbital eccentricty
of the solution}
\label{solve}
\end{figure}
\begin{figure}
\includegraphics[angle=-90,origin=br,width=\columnwidth]{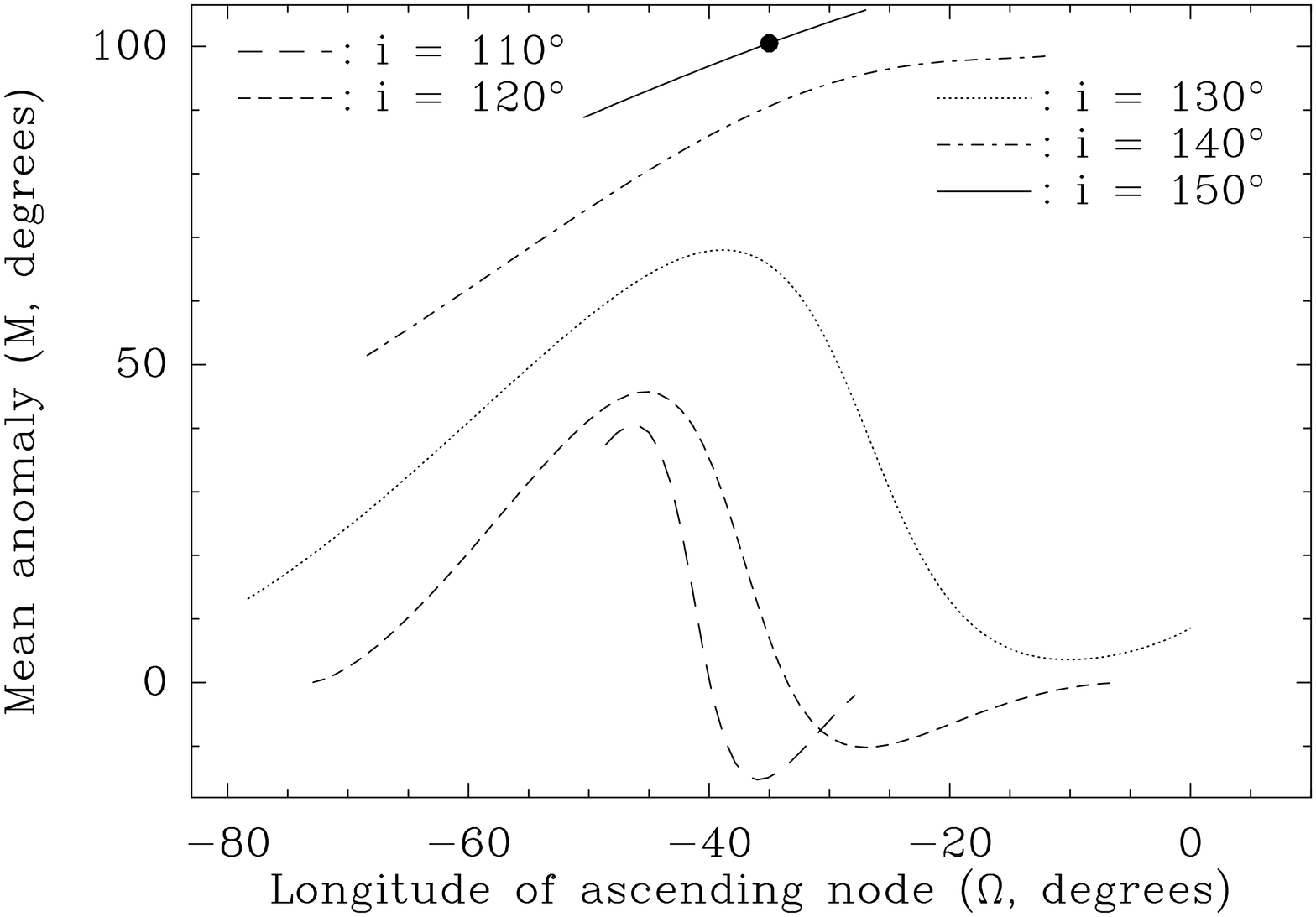}
\caption[]{Same as Fig.~\ref{solva}, but for the present mean anomaly $M$}
\label{solvm}
\end{figure}
\begin{figure}
\includegraphics[angle=-90,origin=br,width=\columnwidth]{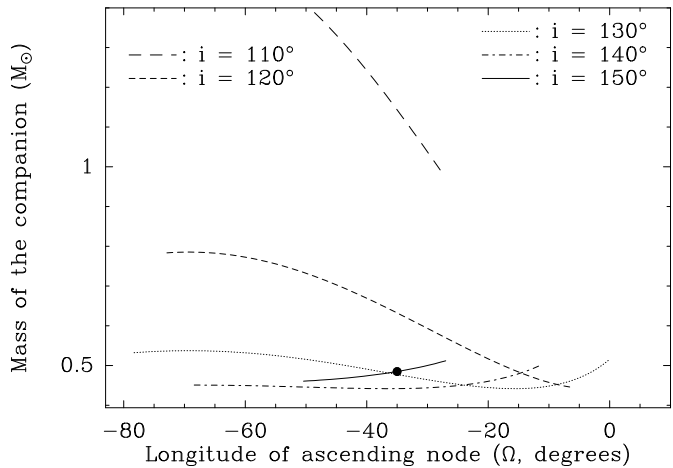}
\caption[]{Same as Fig.~\ref{solva}, but for the fitted 
mass of the \gl~B companion}
\label{solvmass}
\end{figure}
The result of the exploration of the parameter space is shown in
Figs.~\ref{solva}--\ref{solvmass}. Note that contrary to
Figs.~\ref{solm70a}--\ref{solm70m}, solutions are plotted only for
$-83\degr<\Omega<10\degr$; actually there is no convenient solution
out of this range of $\Omega$. We see also that there are solutions
now for $110\degr<i<150\degr$. The orbit is still viewed from south
but it does not exactly lie in the plane of the sky ($i=180\degr$).
Actually with exactly $i=180\degr$, there would be no radial velocity
signature. The significant decrease of the radial velocity as observed
over 25 years forces the inclination $i$ not to be too close to
$i=180\degr$. The solutions are still eccentric, and the present
location of \gl~B is still more or less short after periastron.  The
most interesting outcome concerns the now fitted mass of the companion
(Fig.~\ref{solvmass}). No solution with $m\leq 0.4\,\msun$ is found,
and the more likely solutions correspond to $0.4\,\msun<m<0.6\,\msun$.
This is of course very different from typical brown dwarf values,
but falls well in the range of typical white dwarf masses.

As in the previous section, we display one peculiar solution assumed to
represent a standard solution, charaterized by $i=150\degr$ and
\begin{equation}
\left\{\begin{array}{l}
\dy a=18.42\,\mbox{AU}\;,\quad e=0.3974\;,\quad\Omega=-35\degr\;,\\[\jot]
\dy \omega=18.05\degr\;,\quad M=100.5\degr\;,\quad m=0.4849\,\msun
\end{array}\right.
\label{solpart}
\end{equation}
This solution is marked as a bullet in Figs.~\ref{solva}--\ref{solvmass}.
The orbital period now only 69.7\,yr, and the last
periastron passage occurred in 1983.

In Fig.~\ref{gl86orb}, we show the projection of this solution onto
the plane of the sky like we did it for the orbit corresponding to
Eq.~(\ref{sol1}), and in Fig.~\ref{gl86orbvrad} we show the
correspondant radial velocity curve as a solid grey curve.
The agreement with both the radial
velocity and the astrometric data is very good. Apart from small changes
in the orbital elements, the main difference with the orbit described in
Eq.~(\ref{sol1}) is the mass of the companion. With $m=0.5\,\msun$,
it is obviously not a brown dwarf. 
\section{Discussion}
\subsection{The nature of \gl~B}
From the above analysis, either \gl~B is a brown dwarf,
and then it is unable to explain the RV residuals,
either it is a $\sim 0.5\,\msun$ white dwarf. 
In the former case, another, massive object
is required to explain the RV residuals. In that case,
one should wonder why this object has not been detected yet, unless
it is angularly close to the primary, so that it should disappear
under the coronographic mask used in the images, as suggested by
\cite{els2001}. Given the inclination
we derive for \gl~B, the whole system is thus far from being planar.
Independently from the low probability that such an additional massive
component would be located today in such a position that it could
not be detected, the dynamical stability of the whole system should
be questioned. It is well known \citep{beu97,beu03,kry99} that
multiple systems with high mutual inclinations are often subject to
the Kozai resonance, and that this can lead to instability.

It seems thus more natural to try to attribute the RV residuals
to the sole \gl~B companion. In that case, it must be a
$0.4$--$0.6\,\msun$ object. As from its photometry it cannot be
a main sequence star of that mass, \gl~B turns out to be necessarily
a white dwarf. Our dynamical analysis finally leads to the same
conclusion \cite{mug05} derived from independent spectrophotometric
arguments. 

Based on the present constraint put on the mass of \gl~B and on the
new NACO JHKs photometry, presented in Sect. 2.2, we can now
re-investigate the physical properties of this white dwarf companion,
using predictions of the evolutionary cooling sequences models of
\cite{ber01} for hydrogen- and helium-rich white dwarfs.

\begin{figure*}
\makebox[\textwidth]{
\includegraphics[width=0.9\columnwidth]{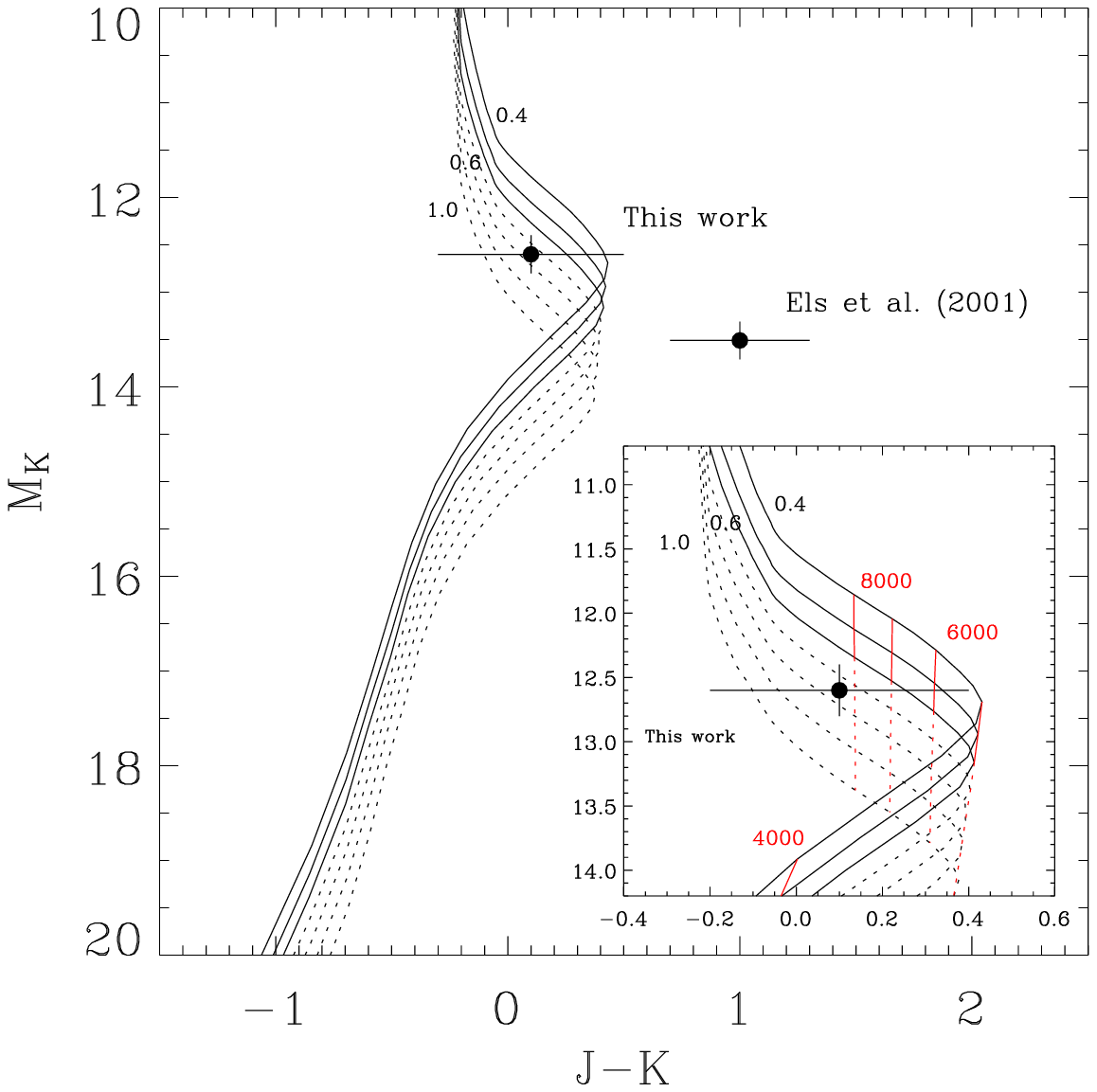}\hfil
\includegraphics[width=0.9\columnwidth]{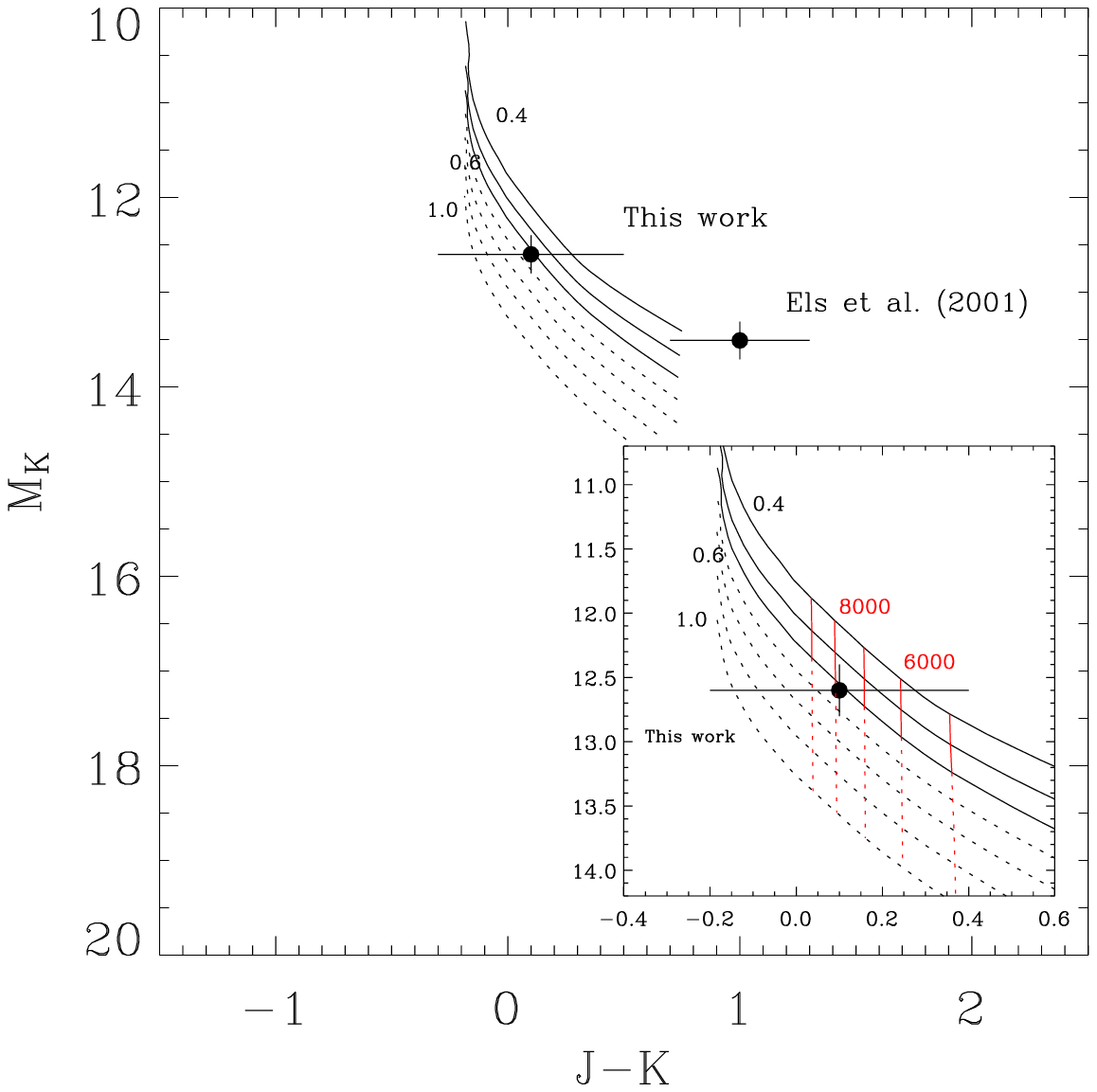}}
\caption[]{Color-magnitude diagram ($\rm{J}-\rm{K}$ vs M$_{K}$) with
model predictions for different masses in two different cases: white
dwarfs with hydrogen-rich (Fig.~\ref{fig:cmd}, \textit{left}) and
helium-rich (Fig.~\ref{fig:cmd}, \textit{right}) atmospheres. The
predictions for a $0.4$, $0.5$ and $0.6\,\msun$ white dwarf, which is
likely the case for \gl~B based on our dynamical analysis, are given
in \textit{solid lines}, the others in \textit{dotted lines}. The
iso-$T_\mathrm{eff}$ lines (\textit{red}) have been also reported in the
zoom-in images. The photometric data from \cite{els2001} and this
work have been reported in both figures for direct comparison with
models predictions.}
\label{fig:cmd}
\end{figure*}

The model predictions are reported in a color-magnitude diagram
($\rm{J}-\rm{K}$ vs M$_{K}$) for both cases: hydrogen-rich
(Fig.~\ref{fig:cmd}, \textit{left}) and helium-rich
(Fig.~\ref{fig:cmd}, \textit{right}) white dwarfs. We can first notice
the discrepancy between the model predictions and the previous
photometric data of \cite{els2001} that \cite{mug05} used to derive an
effective temperature of $5000\pm500$~K for \gl\,B.
Our new NACO photometric data are in very good agreement
with the model and with the dynamical constraints.
Then, if we add the fact that the mass of \gl~B is dyanamically
constrained between ($0.4$--$0.6\,\msun$), we can derive the
effective temperature, the gravity as well as the
cooling age of the
\gl\,B companion based on models predictions. The derived physical
parameters for hydrogen- and helium-rich white dwarf model predictions
are reported in Table~\ref{tab:model}.
\begin{table}[b]
\caption{Physical parameters of \gl\,B based on predictions of the
evolutionary cooling sequences models of \cite{ber01} for
hydrogen- and helium-rich white dwarfs.}
\label{tab:model}
\centering
\begin{tabular*}{\columnwidth}{@{\excs}lllll}     
\hline\hline\noalign{\smallskip} 
Model   & Mass &  $T_\mathrm{eff}$ &  &  Cooling Age \\
        & (M$_{\odot}$)  & (K) &  $\log(g)$ & (Gyr) \\
\noalign{\smallskip} \hline\noalign{\smallskip}
H-rich  & 0.4 & $5500\pm1000$ & $7.66\pm0.02$ & $1.4_{-0.42}^{+1.4}$ \\
\noalign{\smallskip}
& 0.5 & $6000\pm1000$ & $7.86\pm0.01$ & $1.8_{-0.6}^{+1.4}$ \\
\noalign{\smallskip}
& 0.6 & $7000\pm1000$  & $8.01\pm0.01$ & $1.5_{-0.4}^{+0.9}$ \\
\noalign{\smallskip} 
He-rich & 0.4 & $6000\pm1000$ & $7.70\pm0.01$ & $1.6_{-0.6}^{+1.5}$ \\
\noalign{\smallskip}
& 0.5 & $7000\pm1000$ & $7.88\pm0.01$ & $1.3_{-0.34}^{+0.7}$ \\
\noalign{\smallskip}
& 0.6 & $8000\pm1000$ & $8.03\pm0.01$ & $1.2_{-0.29}^{+0.6}$ \\
\noalign{\smallskip} \hline
\end{tabular*}
\end{table}
\subsection{The Initial-Final Mass Relationship}
Both dynamical and spectrophotometric studies come come to the
conclusion that \gl~B is actually a white dwarf.  Let us now
investigate the dynamical implications of this hypothesis. The main
uncertainty concern the initial main-sequence mass of \gl~B before its
evolution to the white dwarf state.  This general problem is known as
the Initial-Final Mass Relationship (IFMR) for white dwarfs
\citep{jef97}. This problem, together with the upper mass limit for
white dwarf progenitors, has been the subject of intense
investigations in the past \citep{wei77,wei87,wei90}.

For what concerns \gl, a first constraint is that in any case, \gl~B
must have been \emph{more massive} than \gl~A in the past
(i.e., $0.8\,\msun$), in order to have more quickly evolved
to the post main-sequence state.

The IFMR is an increasing function of the initial mass. It is usually
measured using white dwarfs that are members of open clusters of known
ages. \cite{wei87} gives a semi-empirical IFMR, but further measurements
of white dwarfs in NGC 2516 \citep{jef97} have shown it was inaccurate.
More relevant relations for various metallicities ($Z$) are given by 
\cite{hur00}. In the following, we will assume the IFMR given by
\cite{hur00} (Fig.~18) for $Z=0.02$. 

Note that this IFMR is different from another one that is sometimes 
shown \citep{iben91,bre93,fag94}, which shows the mass of the 
white dwarf remnant as a function of that of the core at the beginning
of the TP-AGB phase. We are interested into the full initial mass
of \gl~B at Zero Age Main Sequence (ZAMS), so that the first IFMR
is relevant here.
\subsection{Mass loss in a binary system}
Additional constraints can be derived if we consider the past evolution
of the mutual orbit of \gl~A and B. The important post main-sequence mass
loss of \gl~B that led it to its white dwarf state induced an evolution
of the orbit that can be estimated. The general problem of orbital evolution
due to mass loss in a binary system has been theoretically investigated 
by many authors. Basically, one must distinguish between slow and
rapid mass loss. In the former case, the semi-major axis appears
to grow during the mass loss process, while the eccentricity remains
secularly unchanged \citep{jea28,had63,ver72};
in the latter case (rapid mass loss) both the semi-major axis
and the eccentricity grow \citep{bla61,hut81}. A major difference
is that in the case of slow mass loss, the orbit always remains
bound (il just widens), while in the latter case in can be disrupted.
This actually occurs if the mass loss overcomes half of the mass
of the whole system \citep{bla61}. This case corresponds typically
to supernovae.

In the case of \gl, we shall be concerned by the slow mass loss case. The
equations defining the variation of the semi-major axis $a$ and
of the eccentricity $e$  are given by \cite{had63}~:
\begin{eqnarray}
\frac{\rd e}{\rd t} & = & -(e+\cos f)\,\frac{\dot{M}}{M}\qquad;
\label{dedt}\\
aGM(1-e^2) & = & \mbox{constant}\qquad,
\end{eqnarray}
where $M$ is the total mass of the system, $\dot{M}$ the mass loss
rate (due here to \gl~B only) and $f$ is the true anomaly along
the orbit. The second equation arises from the fact that
the specific angular momentum $\vec{C}=\vec{r}\wedge\vec{v}$
is unchanged. The first one is derived assuming that the
change of the specific orbital energy $U$ is only due to the the mass
loss ($\rd U/\rd t= -G\dot{M}/r$) where $r$ is the radius vector
\citep{ver74}. 

If the mass loss is a slow process, Eq.~(\ref{dedt}) can be averaged
over one orbital period. This turns out to give $\overline{\rd e/\rd t}=0$,
which means that the eccentricity is secularly constant
\citep{jea28,had63}. Subsequently, the evolution of the semi-major
axis just obeys the simple rule $aM=\mbox{constant}$. As $M$
decreases, it is obvious that the orbit gets wider. If the total change
of $M$ (only due to \gl~B) is known from the IFMR, it is then
possible to derive the initial semi-major axis.
\subsection{Application to \gl~A and B}
\begin{figure}
\includegraphics[angle=-90,origin=br,width=\columnwidth]{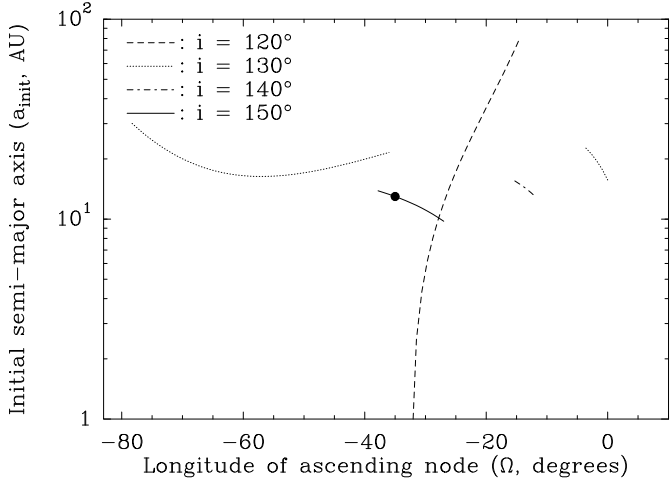}
\caption[]{The initial ZAMS semi-major axis $a_\mathrm{init}$ of the
\gl~B as computed for all solutions displayed in
Figs.~\ref{solva}--\ref{solvmass}, using the IFMR from \cite{hur00}
and $Ma=\mbox{constant}$ . Only the solutions that fit all the 
constraints have been retained (see text).}
\label{solvainit}
\end{figure}
\begin{figure}
\includegraphics[angle=-90,origin=br,width=\columnwidth]{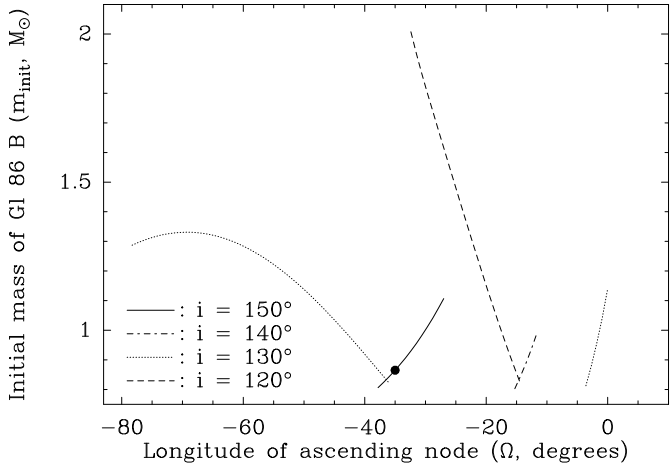}
\caption[]{Same as Fig.~\ref{solvainit}, but for the initial main-sequence
mass of the \gl~B progenitor, assuming \gl~B is presently a white dwarf}
\label{solvmassinit}
\end{figure}
\begin{figure}
\includegraphics[angle=-90,origin=br,width=\columnwidth]{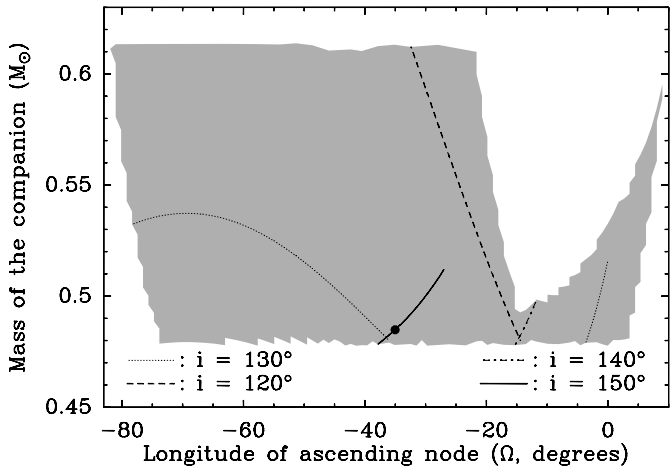}
\caption[]{Same plot as Fig.~\ref{solvmass}, but all solutions
laeding to unphysical or unacceptable values for $a_\mathrm{init}$
$e_\mathrm{init}$ or $m_\mathrm{init}$ have been removed. The
grey shaded area corresponds to all possible values if we let the
inclination $i$ vary.}
\label{solvac}
\end{figure}
If we apply this theory to the case of \gl~B, we are able to derive
the former characteristics of the \gl\ system.  The fit of
Sect.~\ref{fitwd} allows to derive the present day orbital and mass
characteristics of \gl~B ($a$, $e$ and $m$). For each solution,
using the IFMR of \cite{hur00}, we are able to derive the initial mass
$m_\mathrm{init}$, and subsequently the initial 
initial semi-major axis $a_\mathrm{init}$ of the orbit, using
$aM=\mbox{constant}$ . All solutions that lead to
unreaslistic (negative) values for $a_\mathrm{init}$ are
then eliminated; we also eliminate all solutions for which
$m_\mathrm{init}<0.8\,\msun$, as \gl~B must have been initially more
massive than \gl~A. This can be done for every solution that fits the
radial velocity and the astrometric data. This constraint turns out
to be by far the strongest one. 

The result is shown in Figs.~\ref{solvainit}--\ref{solvmassinit}.
In these figures, we plot the resulting values of $a_\mathrm{init}$,
and $m_\mathrm{init}$ for all the solutions displayed
in Figs.~\ref{solva}--\ref{solvmass}. However, we only retain those
solutions which lead to compatible values for $a_\mathrm{init}$,
and to $m_\mathrm{init}>0.8\,\msun$. This is
the reason why the curves are often interrupted. In particular, all
solutions with $i=110\degr$ have been eliminated.

In all cases we have $a_\mathrm{init}<a$
(typically $a_\mathrm{init}\simeq 0.5a$),
showing that the orbit is more detached presently than it was in the past.
This is for instance the case for the solution described in
Eq.~(\ref{solpart}), for which we have
\begin{equation}
a_\mathrm{init}=12.97\,\mbox{AU}\;,\qquad m_\mathrm{init}=0.865\,\msun
\end{equation}
This solution is marked as bullets in
Figs.~\ref{solvainit}--\ref{solvmassinit}. We see that $a_\mathrm{init}$
is not very strongly constrained. The original
mass of \gl~B is better constrained. On Fig.~\ref{solvmassinit},
we see that it may range between 0.8 and $2\,\msun$,
but more probably it was $<1.5\,\msun$. In fact, the solution
giving $m_\mathrm{init}\simeq 2\,\msun$ are those which correspond
to the smallest values for $a_\mathrm{init}$ (see
Figs.~\ref{solvainit}--\ref{solvmassinit}). If $a_\mathrm{init}$ was
too small, the past orbital stability of the exoplanet companion
of \gl~A may be questioned. Obviously this dynamical issue needs to be
investigated into further details. But as a first attempt, let us
consider a possible original configuration of \gl\ with a
$0.8\,\msun$ \gl~A and a $2\,\msun$ \gl~B progenitor. The Hill
radius around \gl~A can thus be estimated to $\sim 0.45\,d$, if
$d$ is the separation between the two stars. If we take for $d$
the periastron of the orbit, with $e\simeq 0.3$ (this is the value
derived for such solutions; see Fig.~\ref{solve}), and if we
assume that the Hill radius must be at least $\sim 2$ times larger
than the 0.11\,AU semi-major axis of the planet to ensure stability,
we derive $a_\mathrm{init}\ga 0.7\,$AU; actually for all solutions with
$a_\mathrm{init}< 1\,$AU, the orbital stability of the exoplanet
is subject to caution.

Another puzzling issue is the way the exoplanet formed. To what extent
was the initial circumstellar
disk of \gl~A that gave birth to its companion truncated by tidal
interaction with \gl~B ?
According to \cite{egg04}, the minimum separation in a binary
that allows a large enough circumstellar disk for planet formation
to survive ranges between 10 and 50\,AU. This could mean that we should
remove all solutions with $a_\mathrm{init}<10\,$AU, which would
result in $m_\mathrm{init}<1.3\,\msun$.

The constraints on  $a_\mathrm{init}$ and
$m_\mathrm{init}$ help to eliminate some of the fitted solutions
in Figs.~\ref{solva}--\ref{solvmass}. This does not change the 
basic constraints on $a$, $e$ and $M$, but refines that on the present
mass $m$ of \gl~B. In Fig.~\ref{solvac}, we show the same plot as in
Fig.~\ref{solvmass}, but all solutions that do not fulfill the constraints
on $a_\mathrm{init}$, $e_\mathrm{init}$ and $m_\mathrm{init}$ have been
removed. In order to explore all possilities, we performed the same
calculation for many inclination values (not only for $i=120\degr$,
$i=130\degr$\ldots). The resulting possibilities are summarized
as grey areas in Fig.~\ref{solvac}. We see that $m$ is fairly well
constrained. It is thus possible to definitively stress that
\begin{equation}
0.48\,\msun\leq m\leq 0.62\,\msun
\end{equation}
and even probably we could say that $m\leq 0.55\,\msun$. The sharp
lower limit at $m=0.48\,\msun$ is due to the lower limit of 
$0.8\,\msun$ for $m_\mathrm{init}$; the upper limit at
$m\simeq0.61\,\msun$ corresponds to $a_\mathrm{init}=0$.
\section{Conclusion}
The identification of the orbital motion of \gl~B around \gl~A,
combined to the measured residuals of the radial velocity data,
allow to severely constrain the whole \gl\ system and its
past evolution. Our dynamical study shows that
\gl~B is very probably a white dwarf, in agreement with the
conclusions of totally independent spectrophotometric study
by \cite{mug05}. The brown dwarf hypothesis of \cite{els2001}
can therefore be definitively ruled out.

The mass of \gl~B is severely constrained by the dynamics. We derive
$0.48\,\msun\le m\le 0.62\,\msun$. The orbit 
is eccentric ($e>0.4$) with a
semi-major axis of a few tens of AU. The associated orbital period is
several hundreds of years at least, and the stars have recently
($5$--$20$ years ago) passed at periastron. The orbit is retrograde
with respect to the plane of the sky, but does not exactly lie in that
plane.  Actually we can say that $120\degr\la i\la150\degr$.

Based on new photometric results on \gl~B and the dynamical mass
constrains, we also re-investigated the physical properties of this
white dwarf companion. Using model predictions of \cite{ber01},
we derived the effective temperature, the gravity and the
cooling age of \gl~B for both hydrogen-rich and helium-rich
atmospheres models of white dwarfs.

When \gl~B was a main sequence star, its mass probably ranged between
$0.8\,\msun$ and $1.5\,\msun$, which implies a spectral type between
K2V and F7V. Its orbit was closer. The strong post-main sequence mass loss
caused the orbit to widen. If it had been
a more massive star, the initial semi-major axis would have been too
small to allow orbital stability for the  exoplanet orbiting \gl~A.

However \cite{saf05} recently used chromospheric index and metallicity
measurements to estimate the age of all known stars harbouring
exoplanets. For \gl~A, they derived an age ranging between 2\,Gyr
and 3\,Gyr. Given the main sequence lifetimes and the white dwarf
cooling times (Table~\ref{tab:model}), assuming this age for \gl~B
would imply that its progenitor had $m_\mathrm{init}\ga2\,\msun$.
This seems to be incompatible with our dynamical constraints.
Obviously, in order to solve this discrepancy, the dynamical evolution
of the whole system, including the exoplanet needs to be investigated
into further details. There are many open questions associated
with this issue: the exoplanet must have survived all the late evolution stages
of \gl~B. If the system is not coplanar,
the exoplanet could have been subject to the Kozai resonance
in the past. Moreover, the planet must have formed in 
a large enough circumstellar disk, which implies a minimum
initial separation of $\sim 10\,$AU.  
All these issues need to be addressed,
and this will be the purpose of forthcoming work.
\end{document}